\documentclass[aip,prb,amsmath,amssymb,twocolumn,reprint]{revtex4}
\usepackage[dvips]{graphicx}
\usepackage{dcolumn}
\usepackage{color}
\usepackage{ulem}
\usepackage{mathptmx, helvet, courier}
\usepackage{bm,amsmath,amssymb}
\usepackage{threeparttable}
\usepackage{multirow}
\usepackage{array}
\usepackage{etoolbox}
\usepackage{mathtools}
\usepackage{hyperref}
\usepackage{adjustbox}
\usepackage{chemformula} 
\usepackage[T1]{fontenc} 
\usepackage{multirow}

\usepackage{titlesec}
\titleformat{\paragraph}
{\normalfont\normalsize\bfseries}{\theparagraph}{1em}{}
\titlespacing*{\paragraph}
{0pt}{3.25ex plus 1ex minus .2ex}{1.5ex plus .2ex}

\usepackage{comment}

\usepackage{tikz}
\usepackage{xcolor}

\begin{document}

\title{Proline isomerization regulates the phase behavior of elastin-like polypeptides in water}

\author{Yani Zhao}
\affiliation{Max Planck Institute for Polymer Research, Ackermannweg 10, 55128 Mainz, Germany}
\author{Kurt Kremer}
\affiliation{Max Planck Institute for Polymer Research, Ackermannweg 10, 55128 Mainz, Germany}
\email{kremer@mpip-mainz.mpg.de}

\begin{abstract}
Responsiveness of polypeptides and polymers in aqueous solution plays an important role in biomedical applications and in designing advanced functional materials. Elastin-like polypeptides (ELPs) are a well-known class of synthetic intrinsically disordered proteins (IDPs), which exhibit a lower critical solution temperature (LCST) in pure water.
The LCST transition can be further tuned by proline isomerization.
Here, we study and compare the influence of {\it cis}/{\it trans} proline isomerization on the collapse of single ELPs in aqueous solution.
Our results reveal that {\it cis} isomers play an important role in tuning the phase behavior of ELPs by hindering peptide-water hydrogen bonding while promoting intramolecular interactions. 
\end{abstract}

\maketitle

\section{Introduction}
Stimulus-triggered intrinsically disordered proteins (IDPs) are involved in a wide range of biological processes. For example, the liquid-liquid phase separation of IDPs is found to contribute to the formation of membraneless organelles~\cite{Hammer2018,Brangwynne:2019}, and the self-assembly of IDPs is associated with numerous human diseases~\cite{Uversky:2008} including neurodegenerative disorders, cancer and amyloidoses.
Synthetic peptide-like polymers that exhibit phase transitions also have broad applications ranging from drug delivery~\cite{Girotti:2016,MacKay:2009,Glassman:2016,Saxena:2015} to polymer materials design~\cite{Stayton:1995,Beer:2014,Mukherji:2020}.
Therefore, the microscopic understanding of the phase behavior of IDPs, which also can be viewed as stimuli responsive polymers, can further provide a suitable platform towards optimized future applications~\cite{Quaroz:2015,Stuart:2010}.

Elastin-like polypeptides (ELPs)~\cite{Quaroz:2015,Roberts:2015} are synthetic IDP-like polymers with pentapeptide repeat sequences Val-Pro-Gly-Xaa-Gly (VPGXG), where the guest residue Xaa can be any amino acid except proline. 
They typically exhibit a lower critical solution temperature (LCST) phase transition~\cite{Hall:2016} in pure water, with a tunable transition temperature $T_l$. It has been shown that the observed $T_l$ depends on the peptide sequences, the chain length, and a number of external stimuli, such as changes in pH~\cite{MacKay:2010}, ion concentration~\cite{Urry:1996}, and pressure~\cite{Tamura:2000}.

Both the simplicity of the amino acid sequence and the stimuli-responsive properties of ELPs make them a suitable platform to study the phase behavior of IDPs and IDP-like polymers at the sequence level. It has been shown experimentally~\cite{Quaroz:2015} that synthetic peptides derived from ELPs which contain Pro-Xaa$_n$-Gly motifs, with $0\le n\le 4$ where Xaa can be any amino acid except Pro and Gly, can be designed to exhibit tunable LCST or UCST (upper critical solution temperature) transitions. In other words, one can synthesize novel biocompatible polymers with tailored phase behavior.   
Additionally, the derived sequence heuristics can also help to predict and identify the phase behavior of existing peptides.

\begin{figure}
\centering
\includegraphics[width=0.4\textwidth]{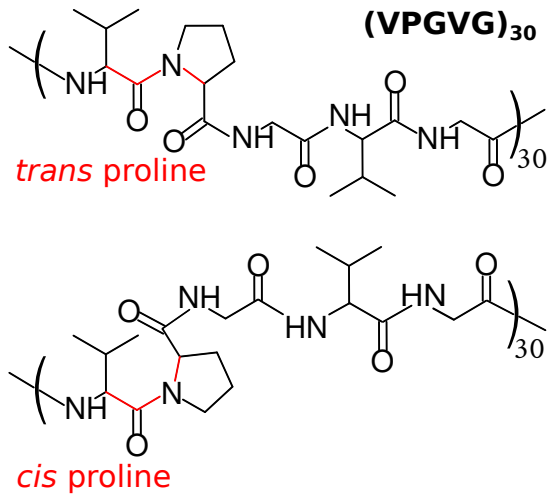}
\caption{Schematic representation of (VPGVG)$_{30}$ with {\it trans} or {\it cis} proline isomers. The backbone residues which restrict the $\omega$ dihedral angle of the Val-Pro amide bonds are marked in red; $\omega=180^\circ$ for the {\it trans} isomer, while it is $0^\circ$ for the {\it cis} isomer. 
} \label{fig:vpgvg}
\end{figure}

ELPs are proline-rich peptides whose phase behavior may be also affected by proline isomerization. 
Proline is the only amino acid with a cyclic side group, i.e., its nitrogen atom is linked to two carbon atoms 
forming a five-membered ring (see Fig.~\ref{fig:vpgvg}). This unique structure stabilizes both {\it cis} and {\it trans} isomers.
While the Gibbs free energy difference between the two proline isomers is only $\sim 2 k_BT$~\cite{Schutkowski:1997,Valiaea:2007}, their transition barrier is rather high, $\sim 30-32$ $k_BT$~\cite{Schutkowski:1997,Zosel:2018}. Therefore, proline isomerization is a fairly slow rate-limiting process~\cite{Keller:1999}, 
which is important in understanding protein folding kinetics~\cite{Brandts:1975}. 
In nature, the {\it trans} isomer is dominant in Xaa-Pro peptide bonds with a {\it trans}:{\it cis} ratio~\cite{Demange:2003} of about 88:12, in excellent agreement with the energy based estimate.  However, one can enhance the {\it cis} isomer content in a number of ways including replacing a proline with a pseudo-proline named $\Psi$Pro~\cite{Keller:1999}, 
using proline isomerase assay~\cite{Schutkowski:1997} or the C(4)-position substituent~\cite{Sonntag:2006}, and also perhaps by ultraviolet photodissociation~\cite{Silzel2020}.

In this work, we study the effects of proline isomerization on the phase behavior of ELPs using all-atom simulations. We consider an ELP sequence of (VPGVG)$_{30}$ with four different {\it cis} proline compositions: (i) all proline residues are in the {\it trans} state $P_{cis} =0$; (ii) half of the proline residues are in the {\it cis} state ($P_{cis}=0.5$), and they are either organized in two blocks {\it {\color{red} cccccccccc ccccc}ttttt tttttttttt}; (iii) or ideally mixed: {\it {\color{red}c}t{\color{red}c}t{\color{red}c}t{\color{red}c}t{\color{red}c}t {\color{red}c}t{\color{red}c}t{\color{red}c}t{\color{red}c}t{\color{red}c}t {\color{red}c}t{\color{red}c}t{\color{red}c}t{\color{red}c}t{\color{red}c}t}; and (iv) all of the proline residues are in the {\it cis} state ($P_{cis}=1.0$).
To simplify the notation, these four cases will be denoted as all {\it trans}, hs-{\it cis}, hm-{\it cis} and all {\it cis}, respectively, in the following text.
Our results show that proline isomerization plays an important role in tuning the phase behavior of ELPs in water. The presence of the {\it cis} isomers facilitates rather compact structures of the peptide. These structures remain largely stable in the temperature range studied, because of enhanced intramolecular and reduced peptide-water hydrogen bonds.
The compactness of the peptide is a function of both the percentage and the position of {\it cis} isomers. The more {\it cis} isomers and the more distributed they are along the sequence, the more compact the chains are. 

\section{Results and discussion}

\begin{figure}
\centering
\includegraphics[width=0.5\textwidth]{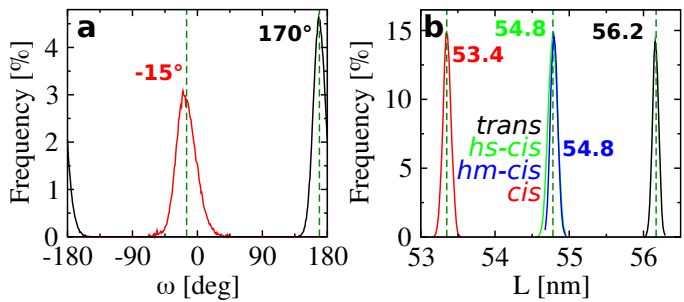}
\caption{(a) The comparison of the $\omega$ dihedral angle between the {\it trans} and {\it cis} isomers at $T=$280 K. The values of $\omega$ at other studied temperatures can be found in Fig.~S1 in the SI. (b) The effective backbone length $L$ of (VPGVG)$_{30}$ in the all {\it trans}, hs-{\it cis}, hm-{\it cis} and all {\it cis} cases. For a given case, $L$ is independent of temperatures.
} \label{fig:omega}
\end{figure}

To characterize the internal structure (backbone orientation) of (VPGVG)$_{30}$ for the four distinct {\it cis} proline compositions considered, we analyze their dihedral angles, $\omega$, of the Val-Pro amide bonds and the effective backbone lengths $L$. The distribution of $\omega$ at $T=280$ K is shown in Fig.~\ref{fig:omega} (a). Note that the $\omega$ distribution remains the same with temperature, see Fig.~S1 in the supporting information (SI).
We find that the average value of $\omega$ is $170^\circ$ for the {\it trans} Val-Pro bonds, while it is $-15^\circ$ for the {\it cis} bonds, roughly consistent the the ideally expected difference of $180^\circ$ between {\it trans} and {\it cis} isomers, respectively. A shift of $10-15^\circ$ in $\omega$ is expected because of the local bending and packing interactions within a molecule. 
In the cases of hs-{\it cis} and hm-{\it cis}, the distribution of $\omega$ in the region with {\it trans} isomers is the same as that from the all {\it trans} case, while that with {\it cis} isomers is the same as the all {\it cis} case.

Fig.~\ref{fig:omega} (b) shows the distribution of $L$ along with the average values $\langle L\rangle=$ 56.2, 54.8, 54.8 and 53.4 nm in the cases of all {\it trans}, hs-{\it cis}, hm-{\it cis} and all {\it cis}, respectively. 
The difference of $\langle L\rangle$ between the all {\it trans} and the all {\it cis} cases is $\Delta  L=\langle L_{trans}\rangle-\langle L_{cis}\rangle=$ 2.8 nm, which gives an average elongation of $\sim 0.1$ nm per proline. 
The elongation has also been observed experimentally~\cite{Valiaea:2007}, which indicates a similar backbone change of {\it cis}-to-{\it trans} isomerization as shown in our simulations.
The results of $L$ and $\omega$ provide a detailed geometric picture of the peptide, i.e., its internal structure with the {\it trans} isomers is different than that with the {\it cis} isomers.

\begin{figure}
\centering
\includegraphics[width=0.5\textwidth]{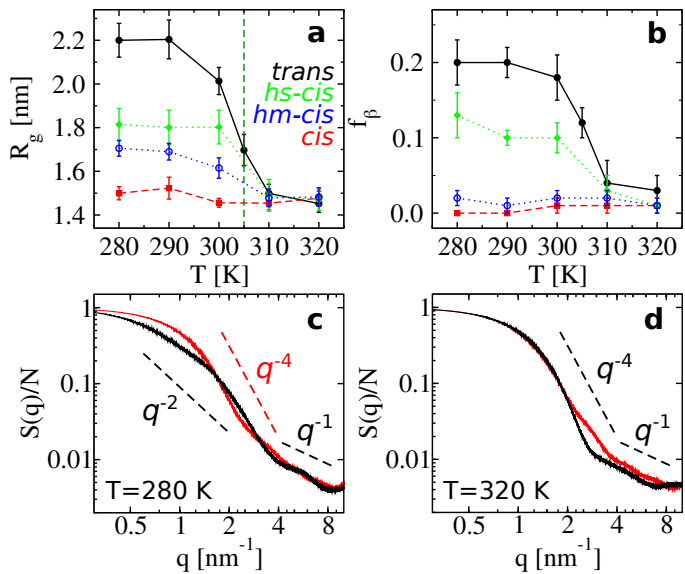}
\caption{(a-b) The results of $R_g$ and $f_\beta$ of (VPGVG)$_{30}$ in the all {\it trans} (black), hs-{\it cis} (green), hm-{\it cis} (blue) and all {\it cis} (red) cases in pure water. The dashed vertical line in panel (a) indicates $T_l\simeq 305$ K detected in our work. (c-d) $S(q)$ at $T=$280 K and 320 K in the all {\it trans} (black) and the all {\it cis} (red) cases.
} \label{fig:purewaterrg}
\end{figure}

In Fig.~\ref{fig:purewaterrg} (a), we show the effects of proline isomerization on the gyration radius $R_g$ of the system as a function of temperature. 
It can be seen from the all {\it trans} data that the peptide shows a well defined coil-to-globule transition upon increase of temperature, the detected LCST transition temperature is around $T_l\simeq 305$ K. This is in very good agreement to experiments, which found $T_l=299$ K for the sequence (VPGVG)$_n$~\cite{Urry:1996}.
At the other extreme, the all {\it cis} case, $R_g$ is nearly independent of temperature, i.e., no LCST-like transition is observed. In the mixed cases, the size of the peptide is in-between the former two. Additionally, the difference between the $R_g$ of hs-{\it cis} and hm-{\it cis} illustrates how the structure of the peptide is affected by the sequence of the {\it cis} isomers along the backbone. In particular, $R_g$ of hs-{\it cis} follows a transition similar to the all {\it trans} case albeit more attenuated (half collapsed and half remains expanded when $T<T_l$), while that of hm-{\it cis} is further declined.
In conclusion, we observe that $T_l$ of the ELP is irrespective of the percentage and position of the {\it cis} isomers, yet its conformational behavior is strongly dependent on the {\it cis} content.

To characterize the local interactions of the peptide, we also estimate the propensity of secondary structure formation. For the sequence (VPGVG)$_{30}$, we expect it to form $\beta-$sheets (it has two valine residues in each pentapeptide) but not $\alpha-$helices (proline and glycine are known to prohibit helix formation). It is indeed the case in the all {\it trans} case but not in the all {\it cis} case, which has also no $\beta-$sheets. Fig.~\ref{fig:purewaterrg} (b) presents the results of $f_\beta$, the fraction of $\beta-$sheets formed by connecting the adjacent $\beta-$strands laterally with hydrogen bonds ($f_\beta$ is obtained by DSSP algorithm~\cite{dssp}).
We find that $f_\beta$ exhibits a similar trend as a function of temperature as $R_g$ in the four systems. In the all {\it trans} case, $f_\beta$ decreases monotonically from $\sim$20\% to $\sim$5\% as $T$ increases from $280$ to $320$ K. The temperature induced decrease in $f_\beta$ is due to the fact that less $\beta-$strands can laterally placed to form $\beta-$sheets as the chain becomes more compact. 
A similar pattern of $f_\beta$ is seen in the hs-{\it cis} case with smaller values for $T<T_l$. A closer look reveals that the decrease is mostly coming from the region with {\it trans} isomers (see Figs.~S2 and S3 in the SI). This observation clearly shows that, the presence of the {\it cis} isomers sterically hinders the formation of hydrogen bonds between local segments, which explains why $f_\beta$ is nearly zero at all temperatures in the hm-{\it cis} and the all {\it cis} cases. 

The global conformation of the ELP was also evaluated by the backbone structure factor $S(q)$. $S(q)$ for the all {\it trans} and all {\it cis} cases at $T<T_l$ ($T=$280 K) and $T>T_l$ ($T=$320 K) are presented in Fig.~\ref{fig:purewaterrg} (c-d). Results for other temperatures are shown in Fig.~S4 in the SI. Our $S(q)$ data show that at $T=280$ K the all {\it cis} chain assumes a globular state ($q^{-4}$ scaling), while the all {\it trans} chain more closely resembles a random walk structure ($q^{-2}$). On smaller scales below 2 {\it nm}, $S(q)$ is very similar in both cases.  At $T = 320$ K the all {\it trans} chains collapse into an even more pronounced globular structure compared to the all {\it cis} case, as demonstrated by Fig.~\ref{fig:purewaterrg} (d) ($S(q)$ of the all {\it cis} chain remains essentially unchanged from $T<T_l$ to $T>T_l$). Data for the hs-{\it cis} and the hm-{\it cis} cases interpolate between these two extremes are shown in the SI. 
The power laws indicated by dashed lines are used to guide the eye.   
Clearly, the peptide behaves roughly similar to a short polymer chain in between the $\Theta$ and the collapsed state in the all {\it trans} case, and it becomes significantly more compact as $P_{cis}$ increases. 
At high $q$ when $q>\frac{2\pi}{l_k}\sim$ 3 nm$^{-1}$, the scaling of $S(q)$ transitions to $q^{-1}$ in all cases, where the peptide behaves essentially like a rigid rod. 
These data agree well with the observations obtained from $R_g$ and $f_\beta$. Moreover, the obtained Kuhn length $l_k\sim 2$ nm (the length of 5 to 6 residues) agrees with our previous simulation~\cite{Zhao:2020a}  and the experimental results~\cite{Schmidt:2010} for ELPs. We also calculated the hydrodynamic radius $R_h$ of (VPGVG)$_{30}$, data are shown in Table~S1 in the SI.
For ideal chains, the value of $R_g$ can be also estimated by the Kuhn length as $\langle R_g^2\rangle=Ll_k/6$. Using $L=56.2$ nm (see Fig.~\ref{fig:omega}) and the value of $R_g$ shown in Fig.~\ref{fig:purewaterrg} results in a by far too small value for $l_k$, revealing significant deviations from a Gaussian structure. In general the data presented here agree quite well with experimental data obtained from dynamic light scattering~\cite{Schmidt:2010} (for details of the different radii we refer to the SI).  
The deviations from the classical polymer picture may be due to the fact that  the ELP chain has a fraction of $\sim 20$\% of $\beta-$sheets in the all {\it trans} case at $T<T_l$. 
Nevertheless, the applied model can successfully catch the LCST transition of the ELP (Fig.~\ref{fig:purewaterrg}) and can distinguish between the {\it cis} and {\it trans} proline states (Fig.~\ref{fig:omega}). 
Above the LCST $S(q)$ displays a $q^{-4}$ scaling in all considered cases, which means that the peptide is collapsed regardless of the {\it cis} content.
Note that, although the shape parameters $S(q)$ and $R_g$ in all considered cases are alike at $T>T_l$, the internal structure of the peptide remains very different as shown in Fig.~\ref{fig:omega}.

\begin{figure}
\centering
\includegraphics[width=0.5\textwidth]{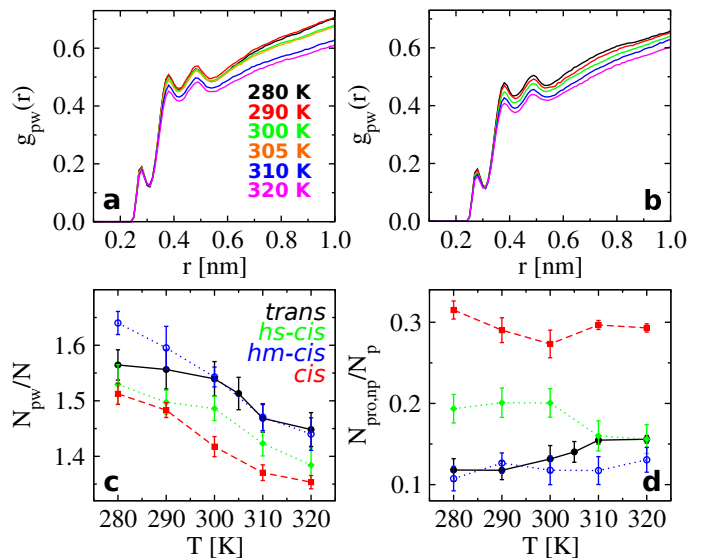}
\caption{(a) The radial distribution function $g_{pw}(r)$ of (VPGVG)$_{30}$ between peptide atoms and water oxygens in the all {\it trans} case as a function of $r$ at different temperatures. (b) The same as (a) but for the all {\it cis} case. 
(c) Normalized H-bonds $N_{pw}/N$ by the total number of residues $N=150$ in the all {\it trans} (black), hs-{\it cis} (green), hm-{\it cis} (blue) and all {\it cis} (red) cases. (d) Normalized intramolecular H-bonds $N_{pro,np}/N_p$ by the number of proline residues $N_p=30$. The color code is the same as panel (c).
} \label{fig:purewaterhb}
\end{figure}

It is known that the LCST phase transition (the coil-to-globule transition upon increase of temperature) is entropy driven~\cite{deGennes1979}, i.e., dominated by the translational entropy gain of the water molecules 
upon collapse of the peptide around $T>T_l$.
The released water molecules can be visualized from the plot of the radial distribution function~\cite{Netz:2017}. As shown in Fig.~\ref{fig:purewaterhb} (a-b), $g_{pw}(r)$ of (VPGVG)$_{30}$ has three peaks within $r\le 1.0$ nm (note that the correlation length of water molecules is less than 2.0 nm in the considered systems~\cite{Mukherji:2020}). We find that the height of the peaks decrease as $T$ increases, because the peptide becomes more compact (see Fig.~\ref{fig:purewaterrg} (a)). However, in the all {\it trans} case a jump of $g_{pw}(r)$ around $T_l$ is observed, while $g_{pw}(r)$ in the all {\it cis} case reduces smoothly. In the mixed cases, $g_{pw}(r)$ of hs-{\it cis} is closer to that in the all {\it trans} case, and $g_{pw}(r)$ of hm-{\it cis} is more similar to that in the all {\it cis} case. 
Interestingly, the amplitude of the three peaks in $g_{pw}(r)$ satisfies $g_{pw,\text{hw-}cis}(r)<g_{pw,\text{all}~trans}(r)<g_{pw,\text{hs-}cis}(r)<g_{pw,\text{all}~cis}(r)$ at $T<T_l$. 
The case of hm-{\it cis} has the largest $g_{pw}(r)$, because it 
has the largest SASA compared with the other cases, see Fig.~S5 (a-b) in the SI. 
Here, SASA measures the surface area of the ELP that is accessible to water molecules.

The results of the normalized peptide water H-bonds $N_{pw}/N$ are shown in Fig.~\ref{fig:purewaterhb} (c), where water molecules can be either hydrogen bond donors or acceptors (see Fig.~S6 in the SI). We find that $N_{pw}/N$ decreases as $T$ increases in all cases, 
which is due to: i) larger thermal fluctuations at higher temperature break the peptide-water H-bonds; ii) the collapse of the peptide with smaller SASA as $T$ increases as mentioned in the previous paragraph.
To see the contribution of the proline residues in the peptide-water H-bonds, we separate $N_{pw}$ into two groups: proline-water and non-proline residues-water H-bonds, the results can be seen in Fig.~S7 in the SI. These data show that the reduction of proline-water H-bonds as $T$ increases is much gentler than that of non-proline residues-water H-bonds. Moreover, less proline-water H-bonds are formed compared with the non-proline residues-water H-bonds. This is because proline can only be a H-bond acceptor, since it doesn't have a N-termini hydrogen atom (see Fig.~\ref{fig:vpgvg}). 
Another observation is that $N_{pw}/N$ shows a plateau at $T<T_l$ and $T>T_l$ but a rapid change around the transition temperature $T_l$ in the all {\it trans} case, and no clear plateau is seen for the hs-{\it cis} case.  On the other hand, $N_{pw}/N$ changes smoothly in the cases of hm-{\it cis} and all {\it cis}. These single chain data indicate that there is no conformational LCST-like transition in the hm-{\it cis} and all {\it cis} cases, and the LCST behavior of the peptide in the hs-{\it cis} case is weakened relative to the all {\it trans} case.

\begin{figure}
\centering
\includegraphics[width=0.3\textwidth]{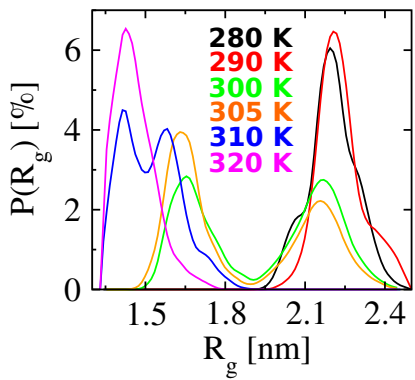}
\caption{The distribution of $R_g$ of (VPGVG)$_{30}$ in the all {\it trans} case at various temperatures. The bimodal distribution of $R_g$ is observed in region 300 K$\le T\le$310 K, at where the LCST transition occurs.
} \label{fig:phasetype}
\end{figure}

We also calculated the intramolecular H-bonds in all considered cases. Fig.~\ref{fig:purewaterhb} (d) shows the results of $N_{pro,np}/N_p$, the H-bonds formed between proline and the other residues of the peptide. We find that $N_{pro,np}/N_p$ in the all {\it cis} case is more than twice as large as in the all {\it trans} case at considered temperatures. The results of hs-{\it cis} is in-between the former two cases and that of hm-{\it cis} is the lowest. Moreover, $N_{pro,np}/N_p$ in the cases of hm-{\it cis} and all {\it cis} remains unchanged with temperature, it decreases in the case of hs-{\it cis}  but increases monotonically in the case of all {\it trans} as $T$ increases from $T<T_l$ to $T>T_l$. These results can be explained by the two competing effects: i) the collapse of the chain results in more intramolecular H-bonds; ii) the breaking of the $\beta-$sheets, if any, leads to fewer intramolecular H-bonds. In the cases of hm-{\it cis} and all {\it cis}, the peptide has nearly no $\beta-$sheets, so the value of $N_{pro,np}/N_p$ is solely dependent on the compactness of the chain. Since the scattered {\it trans} isomers in the hm-{\it cis} case dilute the compactness effects of {\it cis} isomers, its $N_{pro,np}/N_p$ is even smaller than that in the all {\it cis} case. In the hs-{\it cis} case, the number of hydrogen bonds from the half with {\it cis} isomers is barely changing as a function of $T$. In the other half with {\it trans} isomers, there are hydrogen bonds formed to stabilize the $\beta-$sheets at $T<T_l$. This set of hydrogen bonds is gone as $T>T_l$ (see Fig.~S3 in the SI), which caused the reduction of $N_{pro,np}/N_p$ in the collapsed state.
In the all {\it trans} case, $N_{pro,np}/N_p$ slightly increases because the collapse of the chain brings more intramolecular H-bonds at $T>T_l$. Note that there are no H-bonds formed among proline residues. 
Our explanation is also verified by $N_{pp}/N$, see Fig.~S8 in the SI. $N_{pp}/N$ increases monotonically in the cases of hm-{\it cis} and all {\it cis} but decreases in the cases of all {\it trans} and hs-{\it cis} as $T$ increases.
Moreover, the study of a semi-dilute system with several chains indicates that the transition leads to strong chain overlap in the all {\it trans} case while the all {\it cis} chains also seem to aggregate, however with a much weaker tendency to interpenetrate. In other words, the interactions between peptides in the all {\it cis} case appear to be rather weak compared to these in the all {\it trans} case, see Fig.~S9 in the SI.  Whether this holds for much longer chains needs further studies.

Finally, we discuss the type of the LCST transition of (VPGVG)$_{30}$ in the all {\it trans} case. 
Theoretically, both the first-order-like and the second-order-like phase behavior of macromolecules have been observed~\cite{Kremer:1996}, and the standard LCST transition usually is considered to be a $\Theta-$collapse  (second-order-like).
Polyacetals~\cite{Samanta:2016} are examples of the second-order-like LCST transition. Alternatively, in some cases a hysteresis between the heating and cooling procedure around the transition temperature has been observed, which indicates a first-order-like LCST transition. PNIPAm~\cite{Wang:1998} is one such example.
In simulations, a simple way of determining the phase transition type is to check the distribution of the gyration radius $R_g$. A bimodal distribution of $R_g$ near the transition temperature indicates a first-order-like transition, while an unimodal distribution indicates a second-order-like transition. Fig.~\ref{fig:phasetype} presents the $R_g$ distribution of (VPGVG)$_{30}$ at different temperatures. The bimodal distribution around its transition temperature $T_l$ clearly demonstrates a first-order-like transition. 
Note that for such a small peptide, its phase behavior can be also affected by finite-size effects~\cite{Kremer:1996}. Thus a more 
detailed analysis of the transition itself will be a subject of future work.

\section{Conclusions}

We have studied the effects of proline isomerization on the phase behavior of an ELP with sequence (VPGVG)$_{30}$ in water. Our results have shown that proline isomerization plays an important role in tuning the phase behavior of the peptide. In particular, the peptide exhibited a first-order-like LCST transition if all of its proline residues were in the {\it trans} state, while no LCST-like transition has been observed if all prolines were in the {\it cis} state. Moreover, we have found that the number and composition of {\it cis} proline isomers acted cooperatively in determining the global size and the propensity of secondary structure formation of the peptide.
Our work may serve as an inspiration in designing new (bio-)polymeric materials, and opens a novel direction of regulating the phase behavior of ELPs and other proline-rich peptides.

\section*{Acknowledgments}

We thank Debashish Mukherji and Joseph F. Rudzinski for critical reading of the manuscript. Y.Z. thanks Debashish Mukherji for fruitful discussions.
This work has been supported by European Research Council under the European Union's Seventh Framework Programme (FP7/2007-2013)/ERC Grant Agreement No. 340906-MOLPROCOMP.





\bibliography{reference}

\end{document}